# Design and Development of Low Cost Multi-Channel USB Data Acquisition System for the Measurement of Physical Parameters

Nungleppam Monoranjan Singh
Dept of Instrumentation & USIC
Gauhati University
Guwahati – 781014, Assam, India

Kanak Chandra Sarma
Dept of Instrumentation & USIC
Gauhati University
Guwahati – 781014, Assam, India

Nungleppam Gopil Singh
Department of Computer Science and Engineering
Tezpur University
Tezpur – 784028, Assam, India

## ABSTRACT
This paper describes the design and development of low cost USB Data Acquisition System (DAS) for the measurement of physical parameters. Physical parameters such as temperature, humidity, light intensity etc., which are generally slowly varying signals are sensed by respective sensors or integrated sensors and converted into voltages. The DAS is designed using PIC18F4550 microcontroller, communicating with Personal Computer (PC) through USB (Universal Serial Bus). The designed DAS has been tested with the application program developed in Visual Basic, which allows online monitoring in graphical as well as numerical display.

## General Terms
Data Acquisition System (DAS), Universal Serial Bus (USB), Microcontroller.

## Keywords
Data Acquisition System (DAS), temperature, humidity, online monitoring.

## 1. INTRODUCTION
As the computer technology advances, the performance and the availability of the PCs and Laptop become reliable, common and also the prices are falling drastically. Thus, the design and development of the low cost PC based DAS using microcontrollers for use in various fields has been a challenging task. Research is going on in various fields for the design and development of low cost real time DAS [1-5]. Physical parameters such as temperature, pressure, humidity, light intensity etc., are generally slowly varying signals. They can be sensed by respective sensor or transducer giving changes in electrical parameters. In the laboratories or industrial environment, it is very much essential to monitor and/or control such physical parameters. Manual observation and recording of such parameters for continuously for a long time is almost impossible and it cannot fulfill the current requirements in terms of the accuracy and time duration. The efficient solution of this problem is to develope data logger or DAS [6-7]. The present work is to explore the design and development of the low cost multi channel USB DAS using PIC18F4550 microcontroller for continuous monitoring and storing of the physical parameters such as temperature, humidity, etc. Most of the researcher develop application program to customize the readymade data acquisition (DAQ) cards for their specific application. The unique feature of our designed DAS is that, it is designed and developed with commonly available components in the market at low cost; firmware and application program are also developed and are user friendly.

## 2. METHODS AND MATERIALS
The block diagram of the experimental setup is shown in Figure 1. The designed DAS connected with temperature and integrated humidity sensor is shown in Figure 2. The circuit diagram of the designed DAS is shown in figure 3. The hardware, firmware and software description of the system for the real time monitoring of temperature and humidity are described below:

### 2.1 Data Acquisition Unit
A data acquisition system (DAS) has been developed using PIC18F4550 [8], which is a 40/44-Pin, High-Performance, Enhanced Flash, USB Microcontrollers with nano Watt Technology. Figure 1 shows the block diagram of the system so designed. It uses 8 (eight) analog input channels (AN0 through AN07) having 10 bit resolution ADC, in which the entire operation is controlled by the firmware. A B-Type USB socket is used to communicate with the USB port of the PC. The PCB layout of the circuit is designed using DIPTrace and the DAS is fabricated .

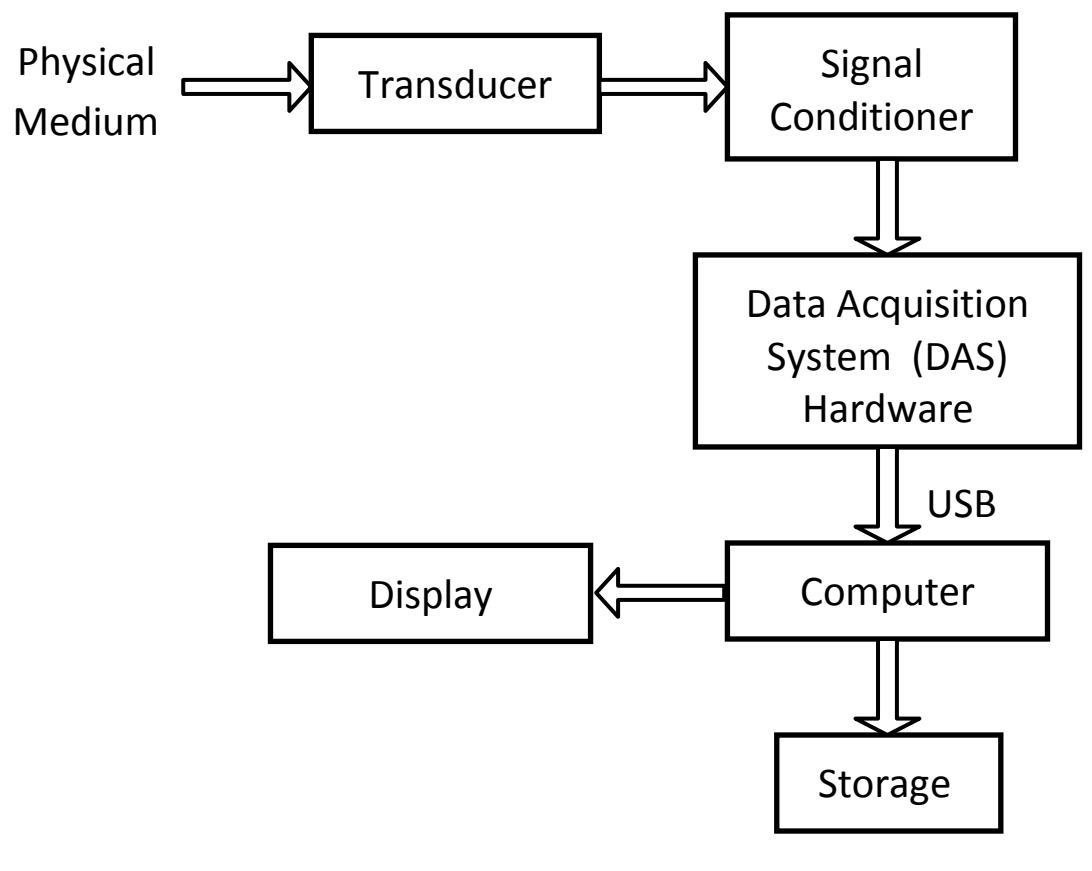

**Figure 1. Block diagram of the Data Acquisition System design**

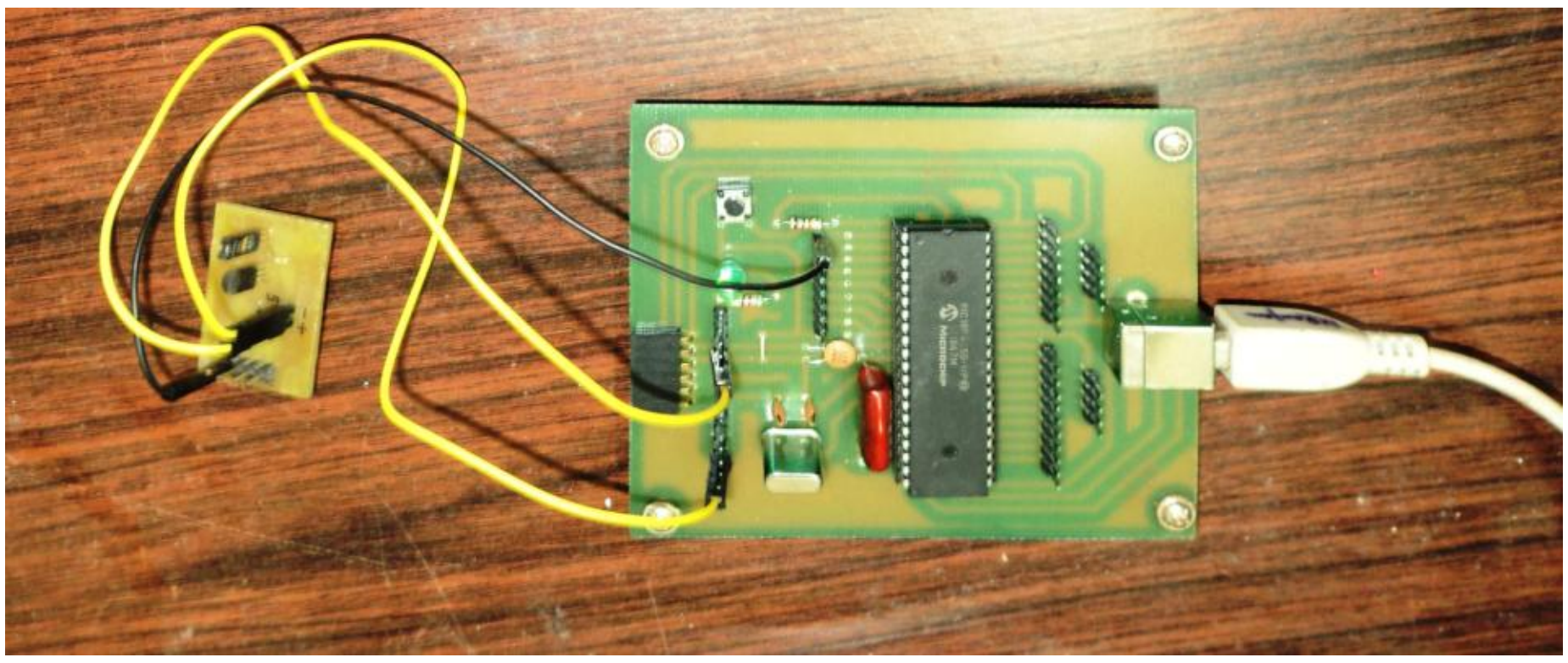

**Figure 2. Low cost multi channel USB Data Acquisition System (DAS)**

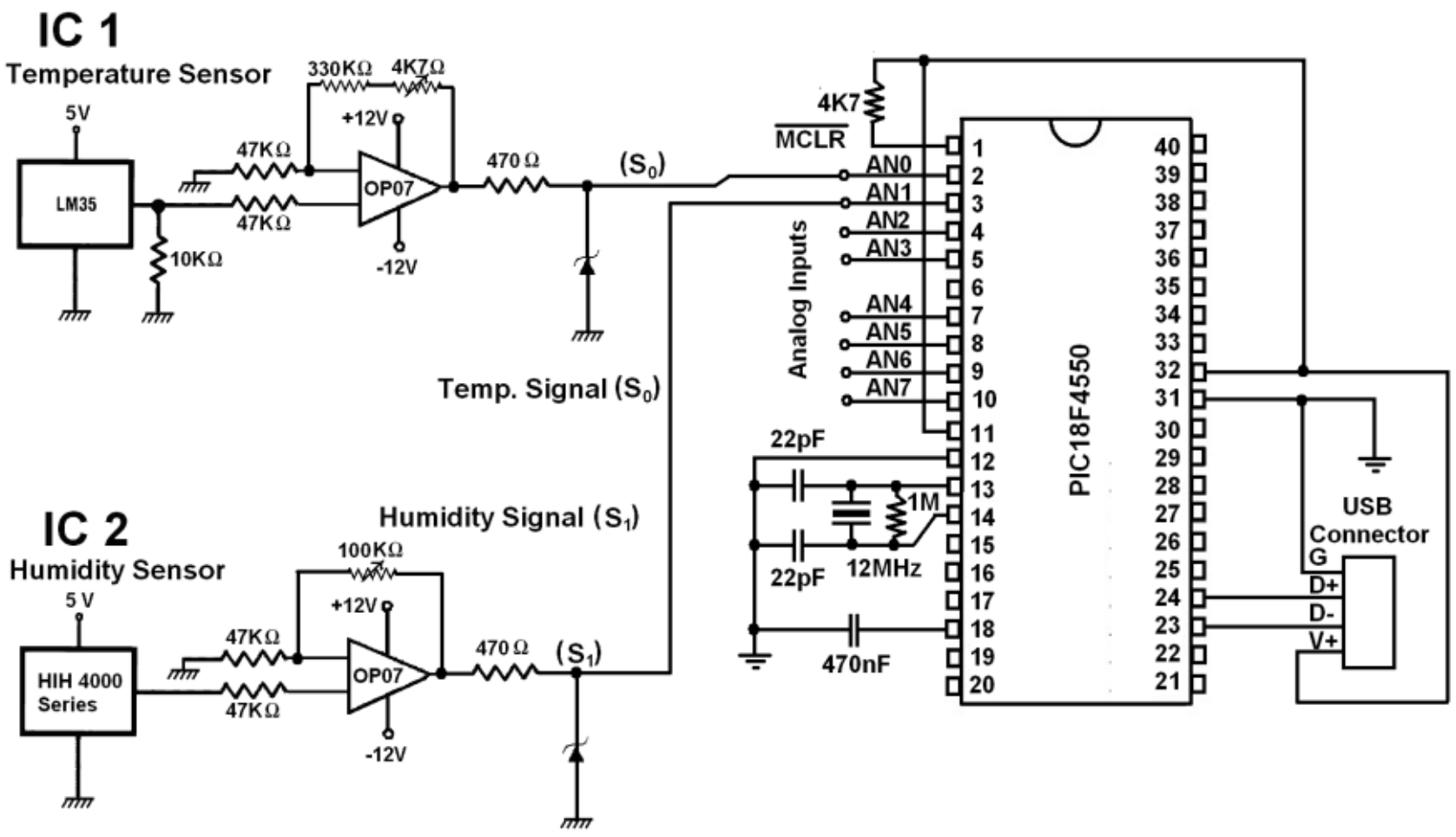

**Figure 3. Circuit diagram of the Data Acquisition System (DAS)**

## 2.2 Sensors

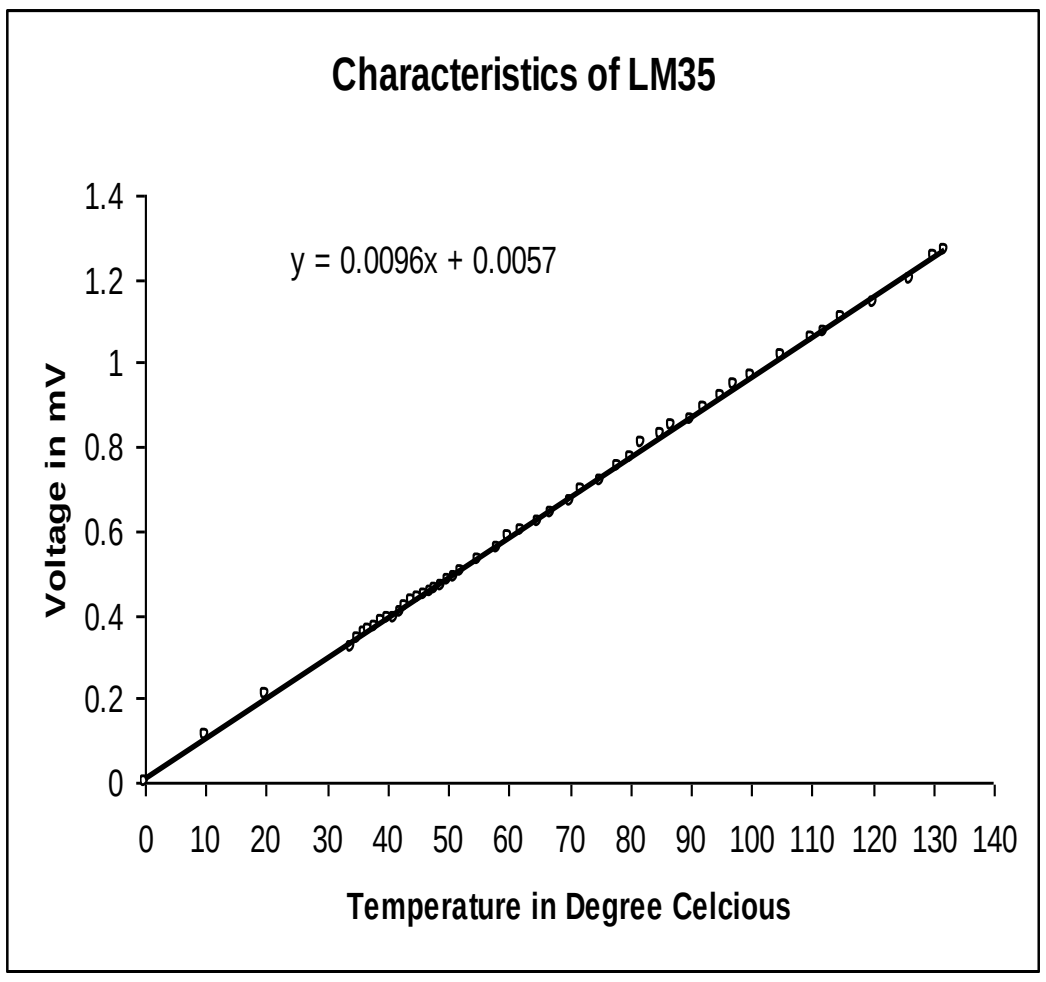

**Figure 4. characteristic graph of the LM35**

An Integrated Circuit (IC) temperature sensor LM35 and an integrated relative humidity sensor HIH 4000 series were used. LM35 is pre calibrated in Degree Celsius. The characteristic graph of the LM35 (Figure 4) from the experimental observation is made for calibration, which is linear [9] and also, the characteristic graph of the humidity sensor from the data sheet is shown (Figure 5) [10].

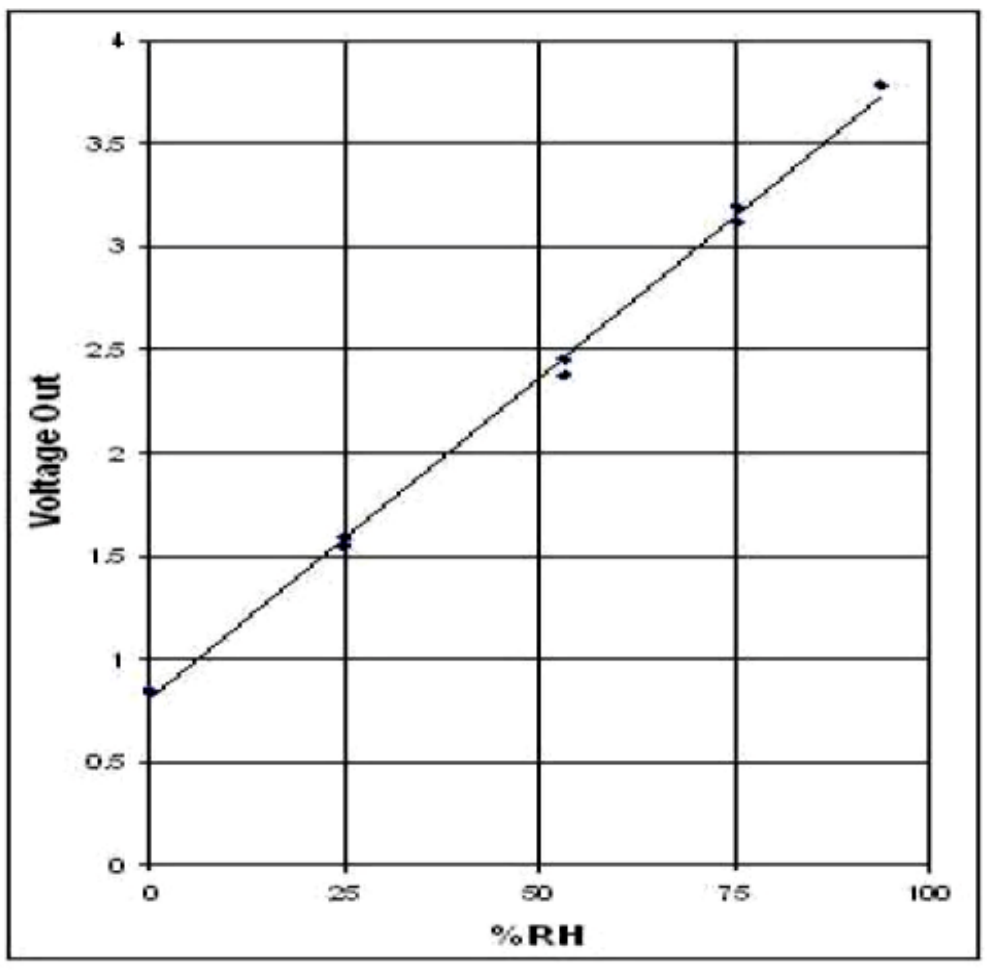

**Figure 5. characteristic graph of the HIH 4000 Sensor**

It uses 8 (eight) analog input channels (AN0 through AN07) having 10 bit resolution ADC, in which the entire operation is controlled by the firmware. A B-Type USB socket is used to communicate with the USB port of the PC. The PCB layout of the circuit is designed using DIPTrace and the DAS is fabricated.

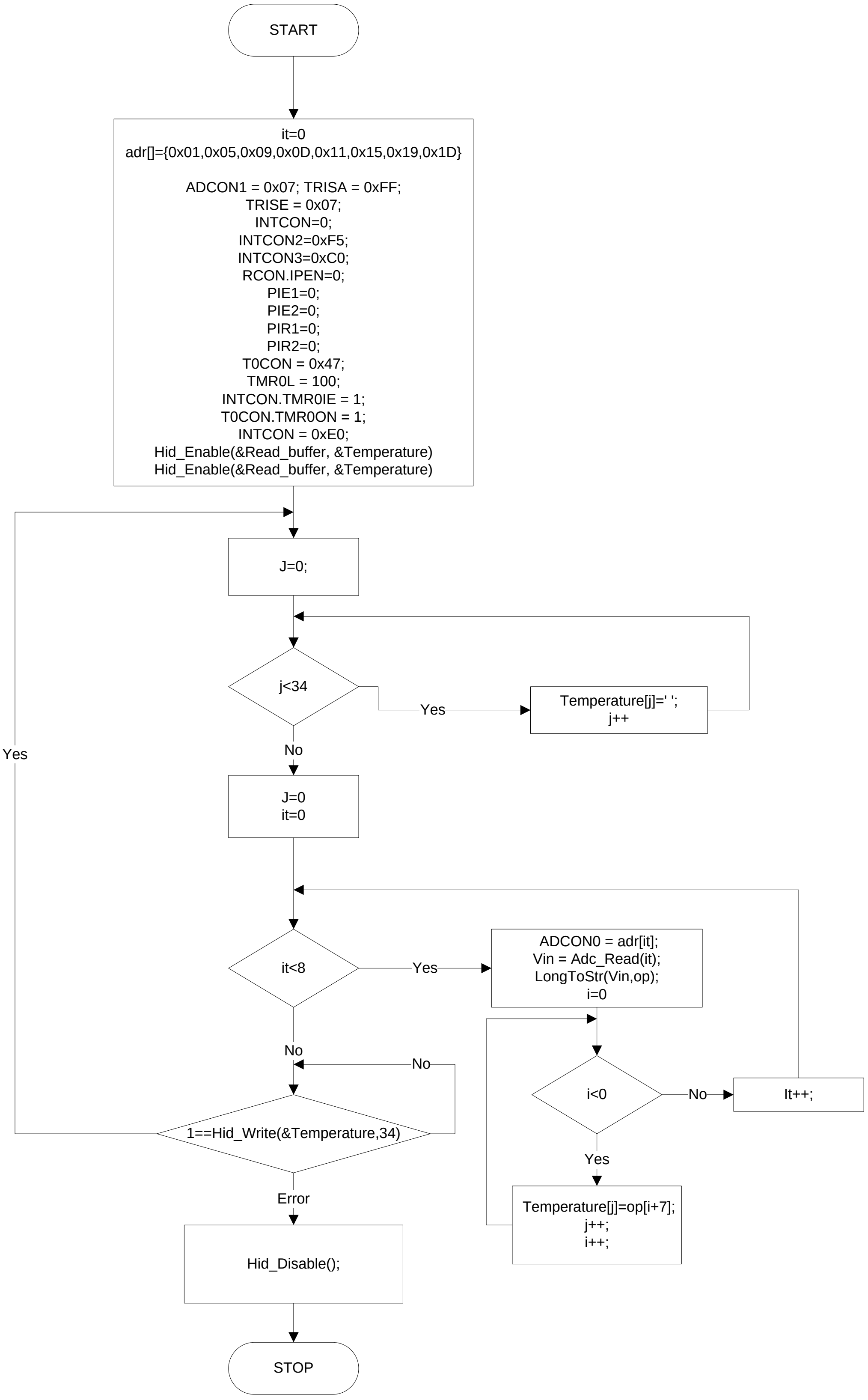

**Figure 6. Flowchart of the firmware for the PIC18F4550 microcontroller.**

## 2.3 Signal Conditioning Circuit

Signals coming out from the temperature sensor IC LM35 and humidity sensor IC HIH 4000 are not suitable for feeding it to the DAS directly. For these, signal conditioning circuits have been designed using OpAmps OP07 and implemented as shown in Figure 3. To protect the over voltage due to accidental or over range, a 5.1 volt zener diode is used in series with a series current limiter resistance. The output is taken across the zener diode (limiting maximum voltage to 5.1 volt) and fed to the input lines of the microcontroller.

## 2.4 Software Section

For the proper functioning of the data acquisition system, a firmware was developed and downloaded to the microcontroller, and an application program was also developed in Visual Basic.

### *2.4.1 Firmware*

A C program was written using MikroC IDE for proper ADC conversion and sending the digitized data to the PC through the USB. The program was compiled and generate a hex file. The hex file so generated was downloaded to the PIC18F4550 microcontroller using PICkit2 programmer. The flowchart of the program so developed is given in Figure 6.

### *2.4.2 Application Program*

For proper acquisition of the data by the PC, graphical display and saving into a file an application program was developed in Visual Basic. For preventing data missed, polling technique was used; that does not require a hardware interrupt. The data so acquired is split into four digit numbers per channel, and displayed. Since it has 10 bit resolution it can read a value from 0 to 5 volt in 1023 steps for a channel. Thus, it has an accuracy of 4.88mV.

For the temperature, the range is set between 0 to 500C. This is converted into voltage from 0 to 5 volt, by the signal conditioning circuit. The range of the relative humidity is from 10 to 90 % RH. Then the signal conditioning circuit converted this voltage range to 1 - 5 volt. The screen shot of the application program is shown in the following Figure 7.

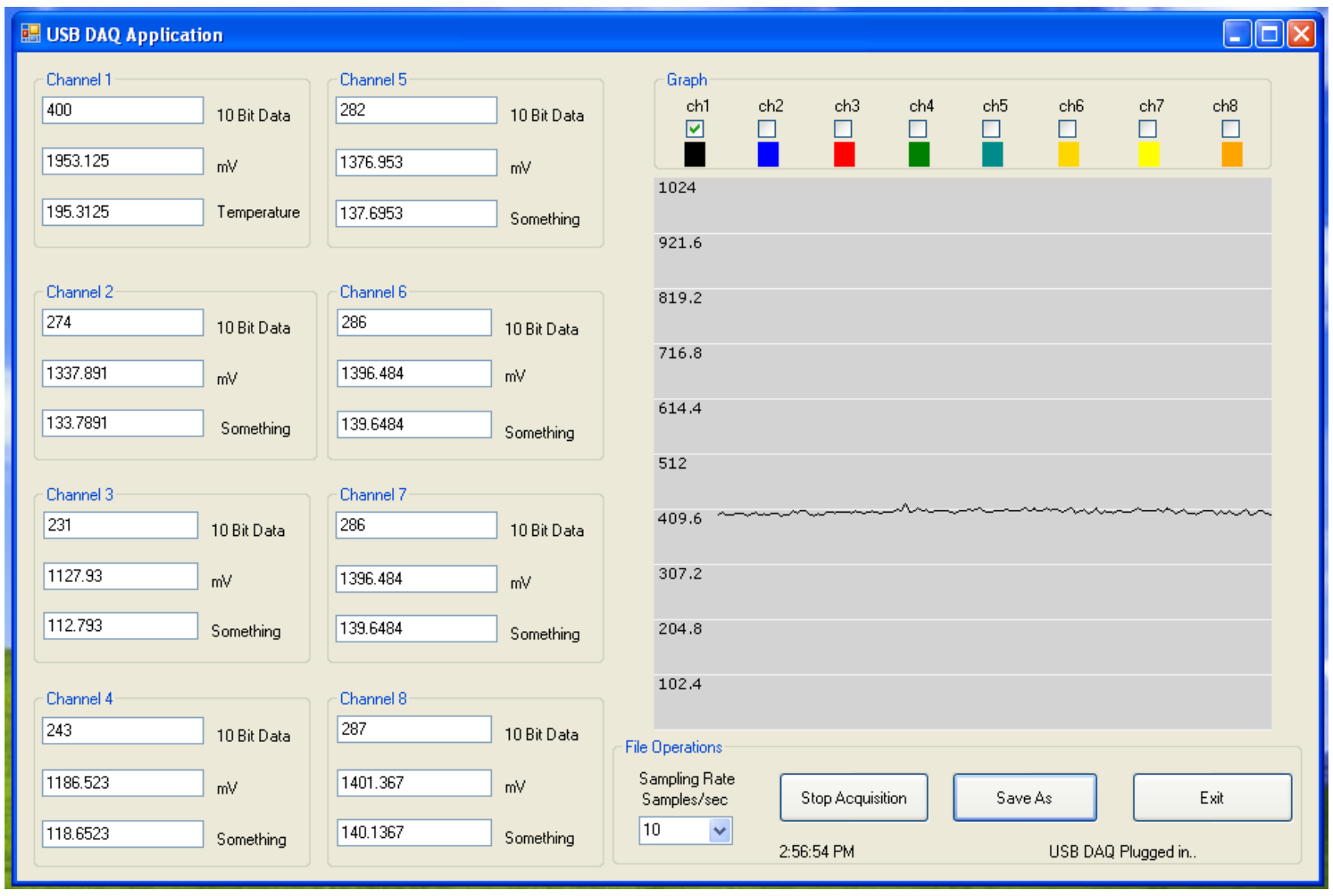

**Figure7. Screen shot of the application program for the designed DAS**

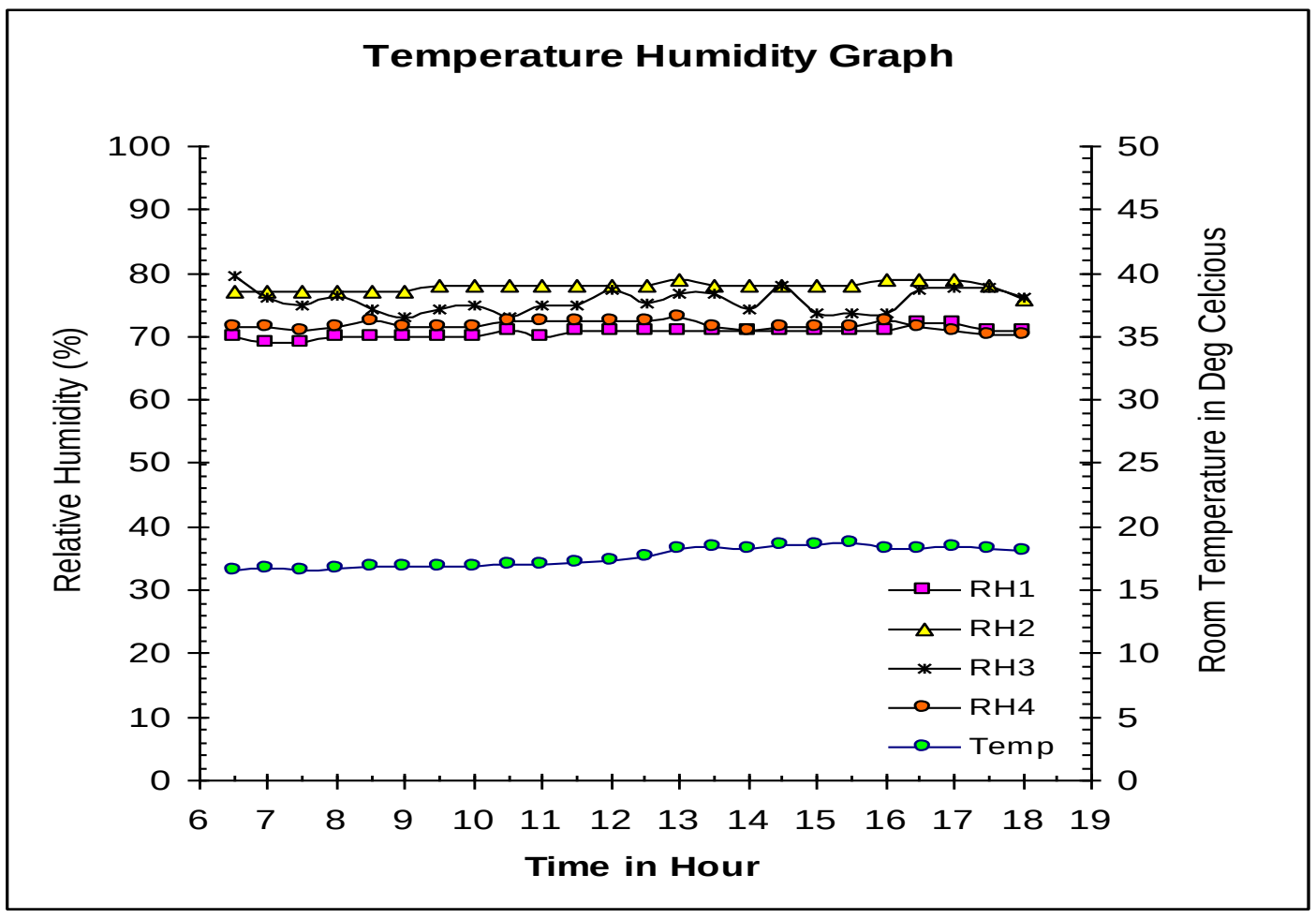

**Figure 8. Reconstructed waveform of the temperature and humidity**

## 3. RESULTS AND CONCLUSION

According to our channel selection a graph is plotted for the selected channel. All the data are stored into the hard disk of the PC into comma separated value (.csv) format. The data so stored can be used for future analysis and also the wave form can be reconstructed. For temperature and humidity one representative curve for 12 hour continuous observation is presented. The obtained real time data for temperature and humidity is presented in the Figure 8. RH1 and RH2 are the observed humidity from the two different digital thermo hygrometers Model No. IT-202, and RH3 is taken from the dry and wet bulb psychometric measurement. RH4 is observed by the designed system. All the four graphs show the similar nature of variations in humidity. The commercial digital hygrometers show a difference of 9% RH as seen from RH1 and RH2. The slight fluctuation in psychometric observation RH3 may be due to fluctuation in air flow and resolution of the thermometers' reading. The variation between RH1 and RH4 is less than ± 2%, where 5% is considerable in general, for RH measurement. Thus, the designed system gives better result

Presently in the designed system, we have used only two channels for giving the inputs for DAS. Since the system is workable for 8 – channels, as tested, some other physical parameters such as light intensity, pressure, displacement, level etc. will also be able to monitor simultaneously using various sensing devices. This system will be useful in research and practical laboratories where acquisition for the measurement, monitoring, analysis and storage of temperature and relative humidity are necessary. In addition, the system can also be used in test and calibration laboratory. The designed system is a low cost with 10 bit resolution having accuracy of 4.88mV (0.0977%) and compatible to PC and laptops. Further, with slight modification, the DAS can also be used for controlling physical parameters.

## 4. ACKNOWLEDGMENTS

The author (N. Monoranjan Singh) acknowledges the financial supports from the Department of Science & Technology, New Delhi under the DST-INSPIRE Fellowship Scheme. The author also acknowledges Microchip for providing help and support in the design and development. The authors also acknowledge the unknown referees for their valuable comments and suggestions for improvement.

## 6. AUTHORS PROFILE

Nungleppam Monoranjan Singh received B.Sc. (Hons.) degree in physics with electronics from Manipur University, Manipur, India in 1999 and the M.Sc. degree in instrumentation from Gauhati University, Assam, India in 2007. He is currently working towards the Ph.D. degree at the Department of Instrumentation and the University Science Instrumentation Centre (USIC), Gauhati University. His areas of research include PC based data-acquisition systems, sensors, instrumentation and control, and embedded systems. He has published papers in national and internal journals. He was awarded INSPIRE Fellow in 2010 by the Department of Science and Technology (DST), New Delhi, India for his research work. He is also life members of the Instrument Society of India (ISOI), International Association of Engineers (IAENG), and Physics Academy of North East (PANE).

Kanak Chandra Sarma received B.Sc. degree and M.Sc. degree in physics from Cotton College, Assam, India, in 1972 and 1974, respectively, and the Ph.D. degree from Guahati University, Assam, India, in 1990. He is a Professor and Head in the Department of Instrumentation and University Science Instrumentation Centre (USIC), Gauhati University. His areas of research include sensors & transducer, instrumentation & control, and thin film nano-materials. He has published papers in national and internal journals. He is reviewer of sensors & actuator, IEEE sensors. He is also life members of the Instrument Society of India (ISOI) and International Association of Engineers (IAENG).

Nungleppam Gopil Singh received B.E. degree in computer technology from Kavikulguru Institute of Technology and Science (K.I.T.S), Maharastra, India in 2010 and M.Tech. degree in information technology from Tezpur University, Assam, India, in 2012. His areas of interest include embedded system, web based applications. He is also life members of the IEEE and International Association of Engineers (IAENG).